\documentclass[%
 reprint,
superscriptaddress,
 amsmath,amssymb,
 aps,
pra,
]{revtex4-2}
\usepackage{amsthm,latexsym,amssymb,amsmath,verbatim}
\usepackage{gensymb}
\usepackage{graphicx}
\usepackage{dcolumn}
\usepackage{hyperref}
\usepackage{chemformula}
\usepackage[T1]{fontenc}
\usepackage{longtable}

\begin{document} 

\title{Photoluminescence excitation spectroscopy of quantum wire-like dislocation states in ZnS}

\author{Alexander Blackston}
\affiliation{These authors contributed equally to this work}
\affiliation
{Department of Materials Science and Engineering, The Ohio State University, Columbus OH 43210, USA} 
\author{Alexandra Fonseca Montenegro} 
\affiliation{These authors contributed equally to this work}
\affiliation
{Department of Materials Science and Engineering, The Ohio State University, Columbus OH 43210, USA} 
\author{Sevim Polat Genlik}
\author{Maryam Ghazisaeidi} 
\affiliation
{Department of Materials Science and Engineering, The Ohio State University, Columbus OH 43210, USA} 
\author{Roberto C. Myers} 
\email{myers.1079@osu.edu} 
\affiliation
{Department of Materials Science and Engineering, The Ohio State University, Columbus OH 43210, USA} 
\affiliation
{Department of Physics, The Ohio State University, Columbus OH 43210, USA} 
\affiliation
{Department of Electrical and Computer Engineering, The Ohio State University, Columbus OH 43210, USA} 

\begin{abstract}
Recent \textit{ab initio} calculations predict 1D dispersive electronic bands confined to the atomic scale cores of dislocations in the wide bandgap (3.84 eV) semiconductor ZnS. We test these predictions by correlating sub-bandgap optical transitions with the density of dislocations formed during strain relaxation in epitaxial ZnS grown on GaP. The densities for four predicted partial dislocations are quantified using scanning electron microscopy-based electron channeling contrast imaging. Room-temperature ellipsometry reveals absorption peaks that scale with dislocation density and align with theoretical predictions. Low-temperature photoluminescence spectra show deep emission peaks matching dislocation 1D band-to-band transitions. Photoluminescence excitation spectroscopy reveals six distinct emission lines with contrasting excitation dependence. Four peaks (2.78, 2.41, 2.20, 1.88 eV), assigned to dislocations, exhibit only modest suppression ($\leq$5$\times$) when excited below the ZnS bandgap, while two other peaks (3.11, 1.53~eV) are strongly quenched ($>$10$\times$). These findings support the existence of efficient, 1D band-to-band radiative transitions within quantum wire-like dislocation core states in ZnS, distinct from typical non-radiative deep-level defects in wide-gap semiconductors.
\end{abstract}

  \maketitle

Dislocations are generally considered detrimental in semiconductors, as they have been associated with nonradiative recombination and degraded transport properties. However, recent theoretical predictions challenge this conventional view, suggesting that certain dislocations can host delocalized, one-dimensional (1D) electronic states with optically active transitions.\cite{polat_genlik_dislocations_2023} Such dislocation bands could serve as intrinsic quantum wires embedded in wide-bandgap materials.

Zincblende ZnS, with its wide bandgap and well-characterized partial dislocations (PDs), is a promising host for exploring these effects. Density functional theory (DFT) calculations predict distinct 1D sub-bandgap states associated with PDs with either Zn- or S-rich core compositions and mixed or pure edge-dislocation structures.\cite{genlik_origin_2024} Validating these predictions experimentally requires correlating dislocation content with optical signatures.

In this work, we examine the optical properties of molecular beam epitaxy (MBE) grown ZnS films on GaP substrates with controlled dislocation densities near the critical thickness.\cite{fonseca_montenegro_log-normal_2024} The dislocation types and densities are quantified via scanning electron microscopy-based electron channeling contrast imaging (SEM-ECCI). Spectroscopic ellipsometry and photoluminescence (PL) measurements reveal sub-bandgap transitions that scale with threading dislocation linear density ($\rho_{TD}$), defined as the total threading dislocation length per unit volume. PL excitation (PLE) spectroscopy further confirms the presence of distinct, radiatively active dislocation states with excitation behavior consistent with dispersive 1D bands. These results establish experimental evidence for quantum wire-like optical transitions in dislocations and suggest a new role for extended defects in semiconductor photonics.

\begin{figure}
    \centering
    \includegraphics[width=3.37in]{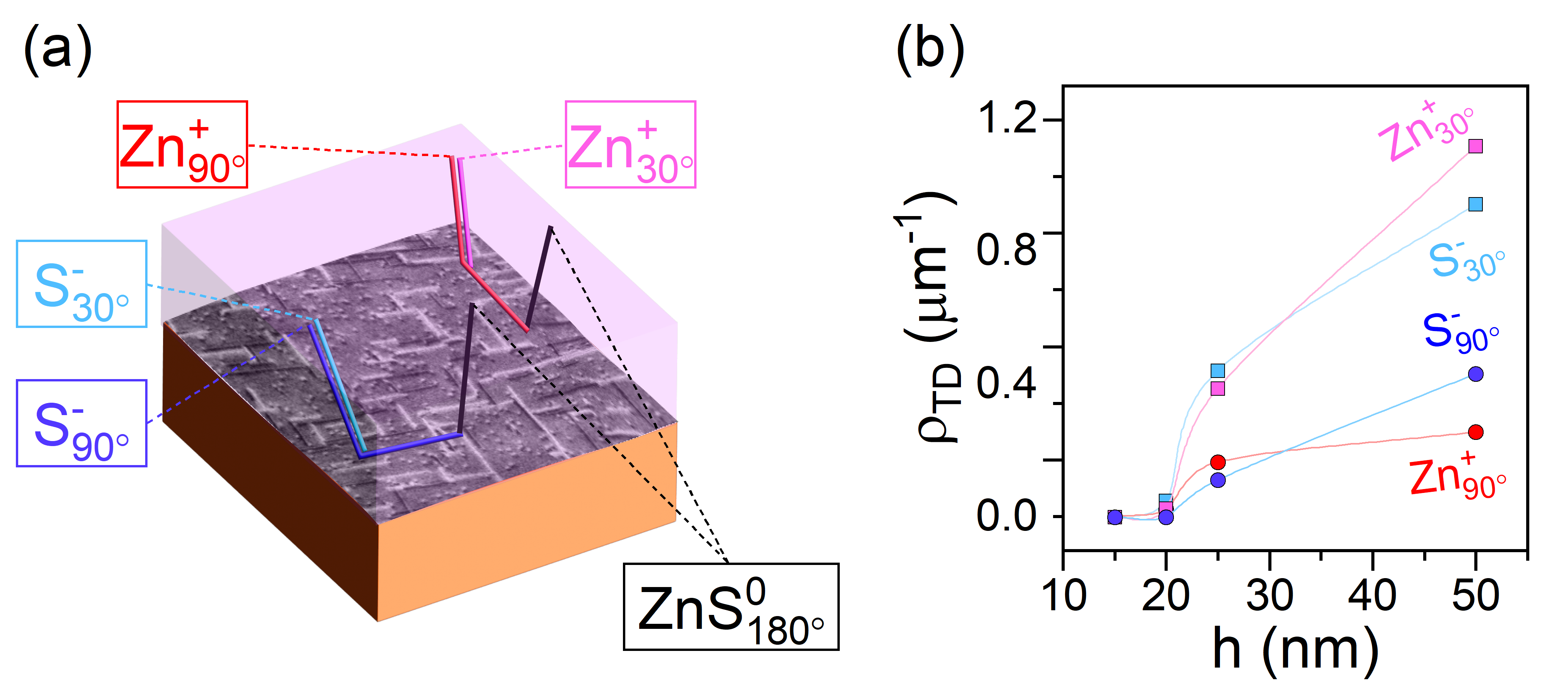}
    \caption{ (a) Schematic of surface terminating threading dislocations (TDs) connected to strain-releiving misfit dislocations (MDs) at the heteroepitaxial ZnS/GaP (001) interface, which are imaged through electron channeling contrast imaging (ECCI),\cite{fonseca_montenegro_log-normal_2024} an example image of which is projected onto the interface. The core structure of MDs and TDs is locked to the line directions, enabling quantification of the densities of Zn-core ($TD_{Zn}^+$), S-core ($TD_{S}^-$), and neutral($TD_{ZnS}^0$). The full dislocations identified by ECCI, dissociate into four possible partial dislocations, with either a Zn-rich or S-rich core, and either pure edge ($90^\circ$) or mixed ($30^\circ$) character. (b) Linear density of TD's ($\rho_{TD}$) as  a function of film thickness ($h$) for the four partials: $Zn_{30^\circ}^+$, $S_{30^\circ}^-$, $Zn_{90^\circ}^+$, and $S_{90^\circ}^-$, see text. Lines guide the eye.}
    \label{fig:densities}
    
\end{figure}

Zinc blende, cubic ZnS, grown on GaP is in biaxial tension due to lattice mismatch ($f=0.57\%$). The strain is relieved by the formation of 60$^\circ$ 1D misfit dislocations (MDs) on \{111\} slip planes with \textbf{a}/2<110> burgers vectors. These line defects form along the heterointerface by nucleation and glide of threading dislocations (TDs). Plan-view SEM-ECCI imaging (e.g. Fig. \ref{fig:densities} (a)) was used to quantify the density of MDs and their TD segments (two per MD).\cite{fonseca_montenegro_log-normal_2024}. Knowing the sign of the strain, and the dislocation line direction, the core composition of the MDs and TDs can be determined, (Zn vs S core).\cite{montenegro_common_2025}. Each interface MD is connected to two surface terminated TD segments, one of which is a neutral 180$^\circ$ pure screw (TD$^0_{ZnS}$, black) and the other a mixed 60$^\circ$ dislocation (TD$^+_{Zn}$ red or TD$^*_{S}$ blue, depending on the line direction).

The full dislocations described above, are expected to dissociate in ZnS due to the low stacking fault energy in ZnS, such that the 60$^\circ$ dissociate into a pair of 30$^\circ$ and 90$^\circ$ (pure edge) dislocations with the same core structure ($Zn^+$ or $S^-$) as the full dislocation (fig. \ref{fig:densities} (a)). Meanwhile the neutral screw TD dissociates into a pair of 30$^\circ$ dislocations, one Zn$^+$  and the other with S$^-$ core. Thus, four types of PD are present ($Zn^+_{30\degree}$, $S^-_{30\degree}$, $Zn^+_{90\degree}$, and $S^-_{90\degree}$), differentiated by their core structure and dislocation angle. Based on previous SEM-ECCI imaging of the ZnS epilayers, $\rho_{TD}$ is estimated and plotted as a function of film thickness ($h$) in Fig. \ref{fig:densities} (b).

\begin{figure}
    \centering
    \includegraphics[width=3.37in]{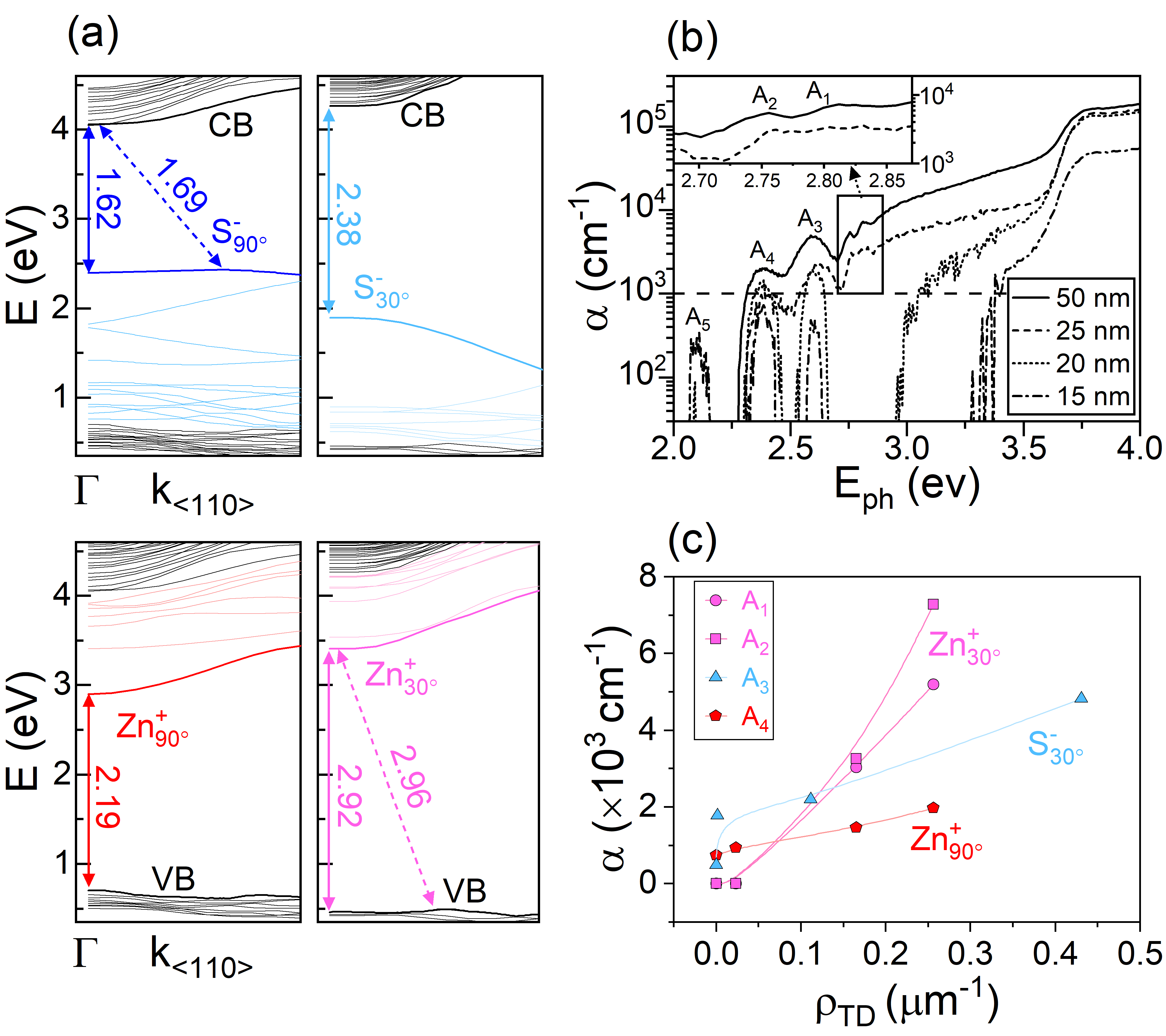} \caption{ DFT predicted dislocation band gap ($E_{gap}^{dln}$) and optical absorption in strain-relaxed ZnS. (a) DFT predicted interband transitions between partial dislocations (PD's) and conduction (CB) and valence band (VB) states.\cite{genlik_origin_2024} The electron energy-momentum (E-k) diagram plotted from the $\Gamma$ point along the PD line direction for the four PDs formed during ZnS epitaxial strain relaxation. Color indicates 1D bands localized at the PD cores ($Zn_{90 \degree}^+$-red, $Zn_{30 \degree}^+$-pink, $S_{90 \degree}^-$-blue, and $S_{30 \degree}^-$-light blue), while black lines are 3D bulk bands. $E_{gap}^{dln}$ energies (eV) for direct (solid) and indirect (dashed) transitions are plotted. (b) Absorption coefficient ($\alpha$) as a function of photon energy ($E_{ph}$) for epitaxial ZnS of various thicknesses. (c) Sub-band gap absorption peak intensity $A_{1-4}$ plotted as a function of the linear threading dislocation density ($\rho_{TD}$), from Fig. \ref{fig:densities}(b), where the PD data are chosen by best matching $E_{gap}^{dln}$ to absorption peak energy ($E_\alpha$).}
    \label{fig:DFT}
\end{figure}

In ZnS, dislocations are predicted by \textit{ab initio} density functional theory (DFT) to produce localized 1D electron gas (1DEG) or hole gas (1DHG) states within the bulk band gap.\cite{genlik_origin_2024} The Zn$^+$ dislocations contain 1DEG bands but negligibly distort the 3D bulk valence band (VB) of the surrounding region. Similarly, the S$^-$ dislocations introduce 1DHGs along the dislocations and minimally distort the bulk conduction band (CB) of the immediate region [Fig. \ref{fig:DFT}a]. The lowest energy k-direct interband transition between the 3D bulk states and the 1D dislocation states, (3D-1D) are plotted in Fig. \ref{fig:DFT}, referred to as the dislocation band gap (E$^{dln}_{gap}$).

The room temperature optical absorption coefficient ($\alpha$) of the ZnS epilayers is measured by spectroscopic ellipsometry as a function of photon energy ($E_{ph}$) and plotted in Fig. \ref{fig:DFT}. As thickness increases and strain-relaxation proceeds, the density of dislocations increases (Fig. \ref{fig:densities})(b)). At the same time, $\alpha$ for  $E_{ph}$ below the band gap  of ZnS ($E_g$) increases. Five sub-band gap absorption peaks, labeled $A_{1-5}$, are observed, as well as a broad band tail. With the exception of $A_5$, the absorption peaks increase with sample thickness and $\rho_{TD}$. $A_1$ and $A_2$ are not observed in the 15 and 20-nm thick samples, but emerge after the plasticity burst, when the MD and TD densities display a rapid increase.\cite{fonseca_montenegro_log-normal_2024} Peaks $A_{1-4}$ are in good agreement with the DFT predicted E$^{dln}_{gap}$, see Table \ref{table}, to within 200 meV, assigning $A_1$ and $A_2$ to $Zn_{30\degree}^+$, $A_3$ to $S_{30\degree}^-$, and $A_4$ to $Zn_{90\degree}^+$. 

\begin{table}[b]
\caption{\label{table}%
Theoretical and experimentally measured optical transition energies in eV for partial dislocations in ZnS. $E_\alpha$ are sub band gap room temperature absorption peaks (Fig. \ref{fig:DFT}) and $E_{PL}$ are the low temperature emission lines (Figs. \ref{fig:pl} and \ref{fig:ple}).
}
\begin{ruledtabular}
\begin{tabular}{lccr}
$E_{gap}^{dln}$ & $E_{\alpha}$ & $E_{PL}$ & $\Delta_{PL-DFT}$\\
\colrule
\colrule

$Zn_{30\degree}^+$ & $A_1$, $A_2$ & $DE_2$ & \\
2.92 & 2.81, 2.75 & 2.79 & -6.1\%\\ 
\colrule
$S_{30\degree}^-$  & $A_3$ & $DE_3$ & \\
2.38 & 2.60 & 2.41 & 1.3\%\\
\colrule
$Zn_{90\degree}^+$  & $A_4$ & $DE_4$ & \\
2.19 & 2.24 & 2.20 & 0.5\%\\ 
\colrule
$S_{90\degree}^-$  &  & $DE_5$ & \\
1.62 & & 1.53 & -5.8\%\\

\end{tabular}
\end{ruledtabular}
\end{table}

\begin{figure}
    \includegraphics[width=3.37in]{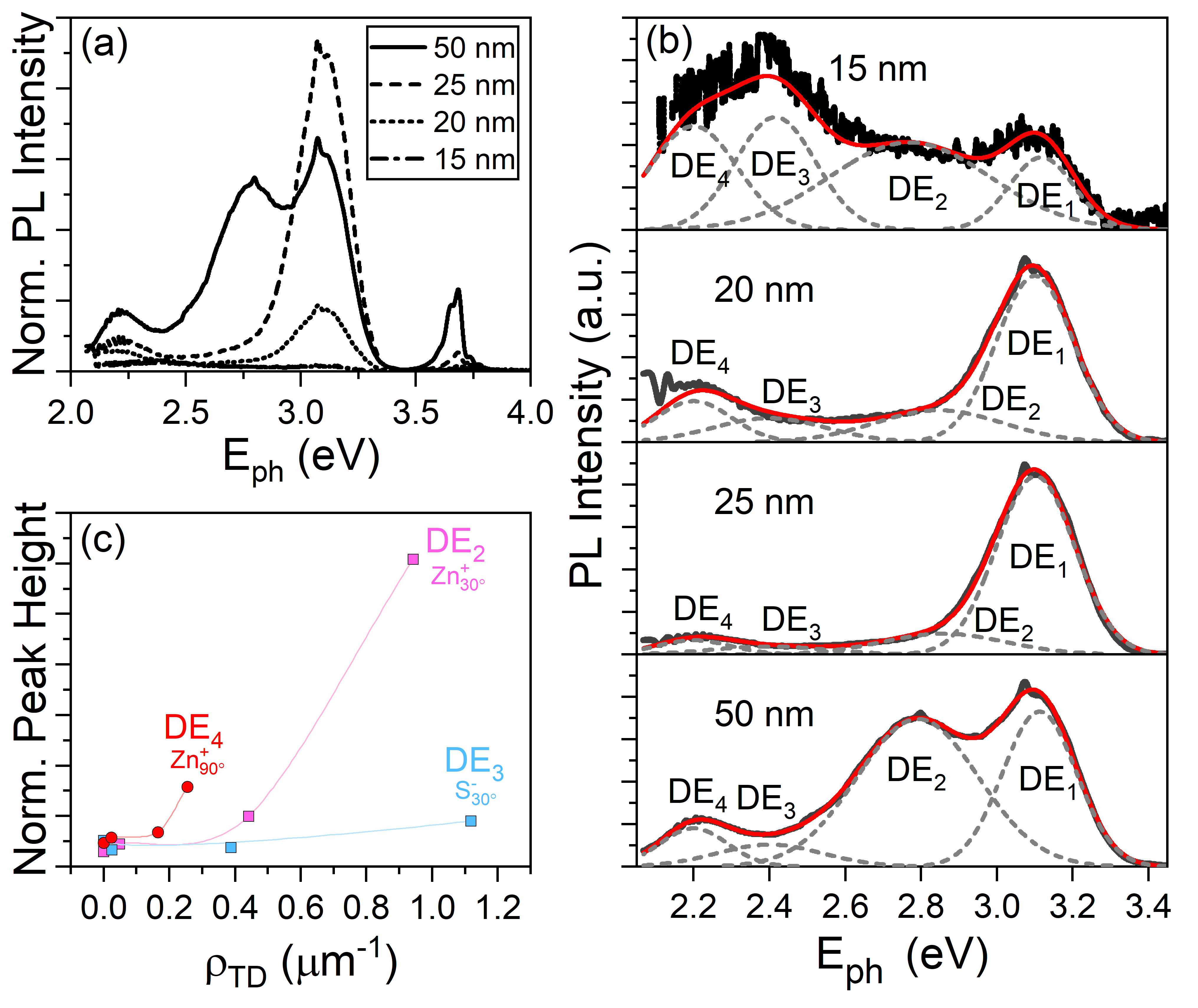}
    \caption{Steady-state photoluminescence spectroscopy of 15 nm, 20 nm, 25 nm, and 50 nm ZnS epilayers.  (a) Thickness dependent PL spectra where PL intensity has been normalized by the fitted ($A_0, X$) peak height (see Methods). (b) PL spectra from part (a) where a Gaussian fit routine has been employed to fit individual PL peaks, labeled $DE_{1-4}$ by decreasing peak energy. (c) Plot of normalized deep level emission peak intensity vs.  ($\rho_{TD}$) for each of the fitted PL peak. Plots are labeled based off of best match between measured PL peak energy and DFT calculated $E_{gap}^{dln}$. }
    \label{fig:pl}
\end{figure}

Low temperature (5 K) PL spectra are collected from each of the ZnS epilayers excitated by 250 nm laser, Fig. \ref{fig:pl}(a). PL peaks consistent with previous reports of band-edge exciton and phonon-assisted transitions in ZnS \cite{ozanyan_near_1996, nam_strain_1998,mitsui_cathodoluminescence_1996} are observed and fit (see Methods). Surface and interface recombination is expected to dominate the photocarrier lifetimes, such that the intensity of PL peaks change strongly with film thickness. To account for this effect, we assume that the near band edge emission, acceptor bound exciton peak ($A_0, X$), is proportional to the steady-state photocarrier density, and normalize the PL spectra by the intensity of the ($A_0, X$) peak. Sharp PL peaks characteristic of GaP are detected near its $E_g$ at 2.4 eV. (see Methods).\cite{dean_intrinsic_1966,dean_absorption_1967, faulkner_excitonic_1969} The GaP related peaks are fitted and subtracted from the spectra to isolate the sub-band-gap emission spectra of the ZnS epilayers. Four broad emission peaks are identified and fit to Gaussian functions centered at 3.11 eV ($DE_1$), 2.79 eV $(DE_2)$, 2.41 eV ($DE_3$),  and 2.20 eV ($DE_4$) (fig. 3(b).) \ref{fig:pl}. Each of these deep level emission peaks are similar to those previously reported of disputed origin.(\cite{lee_optically_1982,mitsui_deep_1998,tsuruoka_origin_2008})\par

The PL peak $DE_1$ is distinct from the others in that its intensity shows an initial increase with $\rho_{TD}$ up until 25 nm, above which it shows a decrease. This would suggest that $DE_1$ does not originate from dislocations. Previous studies have identified an emission peak at $\sim$ 3.1 eV in ZnS that has been ascribed to vacancies.\cite{Cook1992,nam_strain_1998} In contrast, $DE_{2-4}$ are strongly correlated with dislocation content, where their normalized peak heights increase monotonically with $\rho_{TD}$ (Fig. \ref{fig:pl}). A direct linear correlation is not necessarily expected due to competing recombination pathways that change with defect density, e.g. $DE_1$ competing with $DE_{2-4}$.\par

Comparing the PL peak emission energies, we find strong agreement between $A_{1-4}$, $DE_{2-4}$, and the DFT predicted energies (Table \ref{table}). The assignment of $DE_2$ to $Zn^+_{30\degree}$ ($A_1$ and $A_2$), $DE_3$ to $S^-_{30\degree}$ ($A_3$), and $DE_4$ to $Zn^+_{90\degree}$ ($A_4$) is within 200 meV of the DFT. If these PL peaks arose from deep levels, the PL would be red (Stokes) shifted relative to absorption.\cite{boer_optical_2022} Such a red shift is observed for all three peaks, where it is less than 40 meV for most of the assignments, but reaches 190 meV for the peak assigned to $S^-_{30\degree}$ ($DE_3$ and $A_3$).

\begin{figure*}
    \centering
    \includegraphics[width=6.69in]{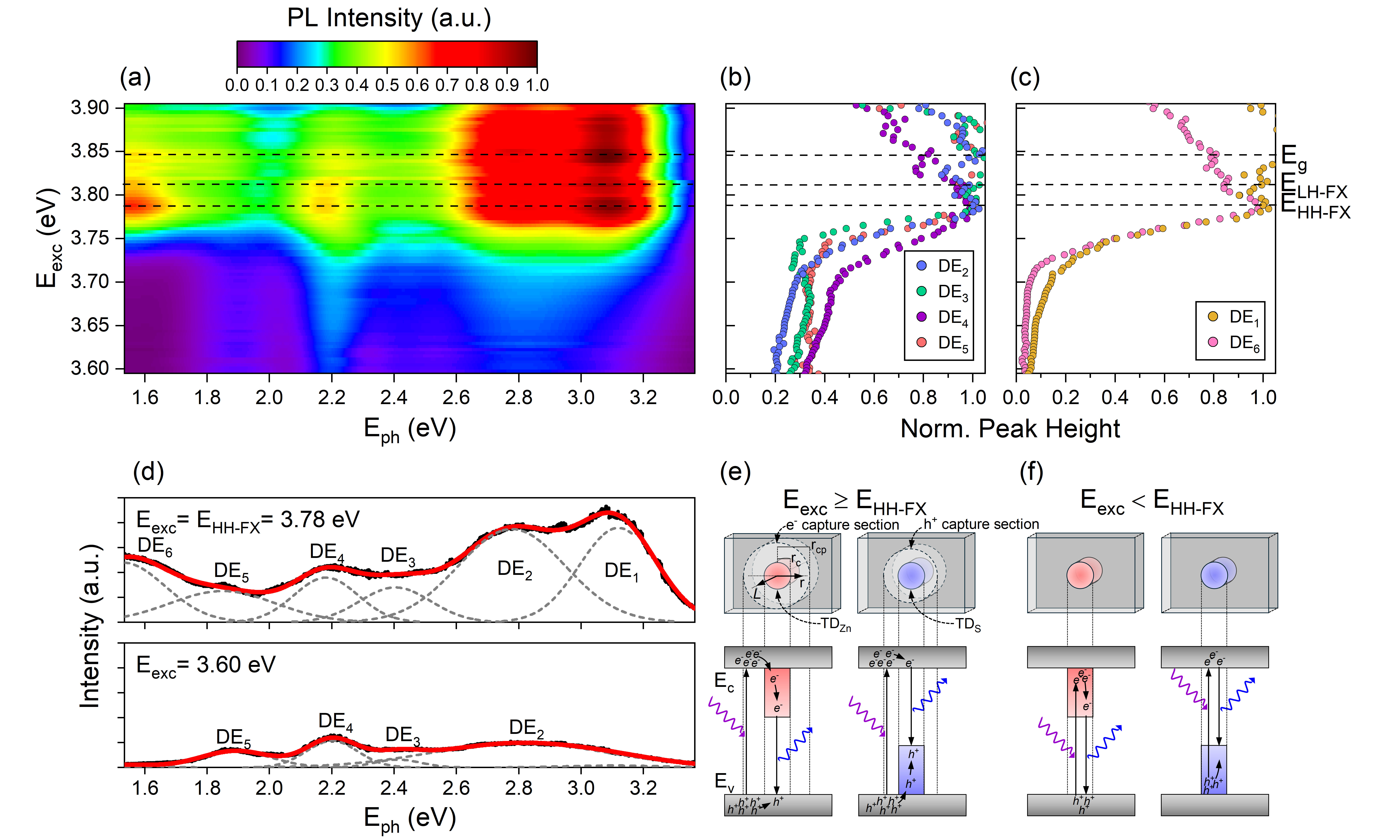}
    \caption{Photoluminescence excitation spectroscopy (PLE) of dislocation-correlated sub-bandgap emission in ZnS. (a) 2D plot of photoluminescence (PL) intensity as a function of photon detection energy ($E_{ph}$) and excitation energy ($E_{exc}$). The fitted PL peak heights are normalized by their value at $E_{HH-FX}$. (b) Fitted peak heights of $DE_2$, $DE_3$, $DE_4$, and $DE_5$, and (c) $DE_1$ and $DE_6$ as a function of ($E_{exc}$). (d) Line cuts with fitted peaks of 2D PLE data in (a), for $E_{exc}$=$E_{HH-FX}$ and $E_{exc}$= 3.60 eV. Schematic real space diagram and energy band diagram displaying absorption and emission processes in dislocations for (e) $E_{exc}\geq E_g$, and (f) $E_{exc}< E_g$.}
    \label{fig:ple}
\end{figure*}

To examine the TD related optical emission/absorption process further, we carry out excitation energy dependent PL spectroscopy (PLE) on the 50 nm sample, with the highest $\rho_{TD}$ and brightest deep level emission in this study. At 5 K, the photon energy of the excitation source ($E_{exc}$) is scanned across $E_g$ of ZnS, from 3.6 eV- 3.9 eV, and PL is detected across 1.5-3.4 eV at each $E_{exc}$, Fig. \ref{fig:ple}(a). Six broad emission peaks are observed for $E_{exc}>E_g$, and are fit to Gaussian functions, Fig. \ref{fig:ple}(d). Peaks $DE_{1-4}$ are observed in all four ZnS samples under 250 nm excitation, as discussed above (Fig. \ref{fig:pl}), with $DE_{2-4}$ being well matched to DFT predicted $E_{gap}^{dln}$ values (Fig. \ref{fig:DFT}), Table \ref{table}. In addition, two additional sub-band-gap peaks are observed, $DE_5$ and $DE_6$, at 1.88 eV and 1.53 eV, respectively. Both are equally close (within 10\%) to the DFT predicted $S^-_{90\degree}$ $E_{gap}^{dln}$ (1.69 eV).

The extracted peak heights for $DE_{1-6}$ are normalized to their value at $E_{exc}=E_{HH-FX}$ and plotted as a function of $E_{exc}$ in Fig. \ref{fig:ple} (b) and (c). For $E_{exc}>E_g$, PLE spectra exhibit multiple peaks centered at $E_{exc}$= 3.788, 3.801, and 3.843  eV, consistent known transitions in bulk ZnS: the heavy hole free exciton ($E_{HH-FX}$), light hole free exciton ($E_{LH-FX}$), and direct band gap transitions, respectively.\cite{tran_photoluminescence_1997}. For $E_{exc}<E_g$, $DE_1$ and $DE_6$ show a $>$10$\times$ decrease in intensity, being undetectable at sufficiently low $E_{exc}$ (Fig. \ref{fig:ple} (c)). In contrast, peaks $DE_{2-5}$ decrease by a smaller factor ($\leq$5$\times$) and remain roughly constant at the lowest $E_{exc}$. 

A TD radiative recombination pathway can explain the PLE spectra. For $E_{exc}\geq E_g$ (fig. \ref{fig:ple}(e), (i) CB electron and VB hole pairs are generated uniformly in the bulk of the ZnS by band-to-band transitions. (ii) The Zn+ (S-) core TDs capture diffusing CB electrons (VB holes) within their diffusion length.\cite{gao_first_2025} The carriers relax to the dislocation band minimum (maximum) through phonon emission,\cite{alkauskas_first-principles_2014} confining them to 1D along the dislocation lines and temporarily forming a 1DEG (1DHG). (iii) VB holes (CB electrons) recombine with 1DEGs in Zn-core TDs (1DHGs in S-core TDs) leading to photon emission at $DE_2$ and $DE_4$ ($DE_3$ and $DE_5$). 

Even though $\rho_{TD}$ is larger for S-core compared to Z-core PDs in this sample (fig. $\ref{fig:densities}$(b)), the PLE for $E_{exc}\geq E_g$ is larger for Zn-core emission ($DE_{2,4}$) than S-core ($DE_{3,5}$), Fig. \ref{fig:ple} (d)). Thus, for the same density of photocarriers, Zn-core TD's capture CB electrons more efficiently than S-core TD's capture VB holes, Fig. \ref{fig:ple}(e). This is consistent with the smaller CB electron effective mass ($m_n$=0.198) than that of VB holes ($m_p=2.526$) in ZnS,\cite{dong_effect_2015} therefore the electron diffusion length ($\lambda_n\propto m_n^{-1/2}$) is greater than that of holes ($\lambda_p$) enabling enhanced carrier capture radius ($r_{cp}$). From the Einstein relations, assuming equivalent scattering rates, the ratio of the capture rates for PDs of different core stoichiometry, but identical structure is, $r_{cp}^{Zn^+}/r_{cp}^{S^-}=\sqrt{m_h/m_n}\approx 3.6$. This roughly accounts for the ratio of PL $DE_2/DE_4\approx 2.10$ in Fig. \ref{fig:ple}(d) but is less consistent with $DE_3/DE_5\approx 1.13$, indicating the necessity of more complex capture models, than simple flat band diffusion. For $E_{exc}$ < $E_g$ (Fig. \ref{fig:ple}(f)), the PLE reduces for all PL peaks as bulk band-to-band transitions become forbidden and only transitions from bulk VB to defect levels or defect to CB are possible and are limited by the density of such defects. In the case of dislocations, optical absorption now occurs directly in the dispersive core bands states, resonantly producing 1D confined electrons (holes) in the $Zn^+$-core ($S^-$-core) TD's, which radiatively recombine and emit PL. This resonant excitation does not require carrier capture (ii) making it more radiatively efficient than $E_{exc}\geq E_g$, with PLE now limited by the overall density of emitters.

Dispersing 1D dislocation bands with a quasi-continuum of states, enable broad excitation spanning well below $E_g$. This is consistent with the PLE spectra of $DE_{2-5}$ (Fig. \ref{fig:ple}(b)). The remaining $DE_{1,6}$ might be due to deep level emission, although their possible dislocation origin cannot be ruled out in this study. As described above, the $\rho_{TD}$ dependence of $DE_1$ observed in Fig. \ref{fig:pl} indicates that it likely arises from defects. Both $DE_1$ and $DE_6$ show rapid quenching for $E_{exc}<E_g$, indicating a distinct recombination process from $DE_{2-5}$. For deep level defects, for $E_{exc}<E_g$, photoionization will arise, with a characteristic absorption edge.\cite{boer_optical_2022} Deep level photoionization is limited by the density of defects, but its radiative emission efficiency is typically several orders of magnitude lower than the competing non-radiative pathways.\cite{dreyer_radiative_2020,zhao_carbon_2025}

This study compares optical emission and absorption spectra of ZnS with quantified dislocation content, with DFT predictions of interband transitions within dislocation cores. Four room temperature sub-band-gap absorption peaks are observed with rough agreement to the DFT, and a strong correlation with $\rho_{TD}$. Low temperature PL also reveals three emission peaks correlated with $\rho_{TD}$. PLE spectra are obtained for six different sub-band-gap emission peaks in epitaxial ZnS with the largest $\rho_{TD}$ in the study. Two of the peaks are only observed when exciting above $E_g$, whereas four of the peaks can be efficiently excited well below $E_g$, indicative of efficient photocarrier excitation and emission within PD cores in ZnS.

\section{\label{sec:level1}Methods}

\subsection{\label{sec:level2}Photoluminescence Spectroscopy}
Spectrally resolved measurements of photoluminescence were taken in the reflection geometry using a tunable Ti-sapphire laser (Coherent Chameleon Ultra II) in conjunction with a SHG and THG generation system (Coherent Vue) as the excitation source. The laser power was monitored using a photodiode power sensor and was tuned to 1 mW using a linear polarizer on a rotation stage. The beam was focused using a 150 mm focal length achromatic lens. The reflected beam and photoluminescence was sent through a long pass filter to filter out elastically scattered light and then analyzed using a 1200 g/mm, 500 nm blaze reflective diffraction grating and Teledyne Vision Solutions Pixis 400 CCD camera housed inside of a Princeton Instruments Acton Series SP-2500i spectrometer. All photoluminescence measurements were carried out in a closed-loop helium cryostat (Montana Instruments Cryostation S50) at 5 K.  

\subsection{\label{sec:level2}ZnS Band Edge Photoluminescence}

\begin{figure}
    \centering
    \includegraphics[width=3in]{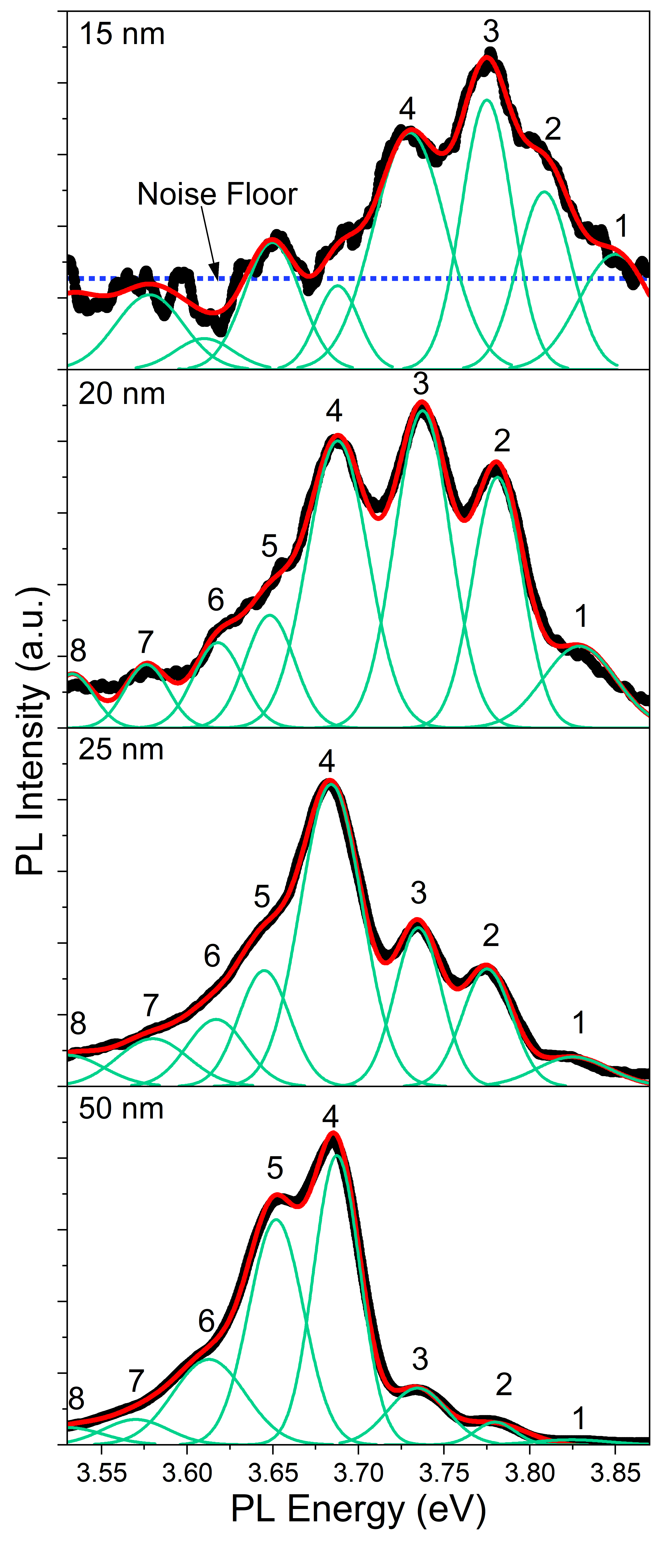}
    \caption{Thickness dependent band edge photoluminescence spectra fit with Gaussian functions. The noise floor is marked for the 15 nm sample spectrum which makes it difficult to conclusively identify any of the phonon replicas of the $(e, A)$ transition peak. 
    \label{supfig1}
    }
\end{figure}

Band edge photoluminescence from ZnS epilayers (t= 15nm, 20nm, 25nm, and 50 nm) are collected with a 250 nm (4.96 eV) excitation source. 
Spectra are fit using Gaussian curves (Fig. \ref{supfig1}). Observed peak energies are compared to previous reports in the literature \cite{ lee_optically_1982, tran_photoluminescence_1997, nam_strain_1998,} (Table \ref{suptable1}). Such studies published photoluminescence spectra that included peaks at 3.826, 3.782 eV, 3.744 eV, and 3.67 eV that are thought to be associated with a neutral donor bound transition $(D^0, X)$, heavy-hole free exciton transition (HH- XF), neutral acceptor bound exciton transition $(A^0, X)$, and conduction electron to acceptor transition$(e, A)$, respectively. Multiple longitudinal optical (LO) phonon replicas of the $(e, A)$ transition have also been observed. Energies consistent with these transitions are found in our spectra. Nam et al. claim that P that has diffused from the GaP substrate into ZnS is acting as the acceptor in the transitions they observed. We obtained comparatively brighter $(e, A)$ peaks form our epilayers which are much thinner then their epilayers and would therefore likely contain a higher concentration of P that has diffused from the substrate, thus supporting this hypothesis.

\begin{table*}
\centering
\begin{tabular}{|l||l|l|l|l|}
 \hline
 \multicolumn{5}{|c|}{ZnS Band Edge Transitions} \\
 \hline
 Peak Num. & Transition & Energy (eV) – This Study & Energy (eV) – Nam et al. & Energy (eV) – Leo et al. \\
 \hline

 1 &$(D^0, X)$ & 3.826 & N/A& N/A\\
 
 2 &$HH-FX$ & 3.780 & 3.782 & 3.790\\
 
 3 &$(A^0, X)$& 3.737 & 3.744 & 3.737\\ 
 
 4 &$(e,A^0)$& 3.688 & 3.664 & 3.667\\
 
 5 &$(e,A^0)+LO$& 3.652 & 3.621 & 3.624\\
 
 6 &$(e,A^0)+2LO$& 3.613& N/A & N/A \\
 
 7 &$(e,A^0)+3LO$& 3.568& N/A & N/A\\

 8 &$(e,A^0)+4LO$& 3.530& N/A & N/A\\
 
 \hline
\end{tabular}
\caption{Summary of measured ZnS transition peaks and comparison to previous literature.
\label{suptable1}
}

\end{table*}

Additionally, a significant increase in intensity of the exciton transition peaks, $(D^0, X)$, $(HH-FX)$ and$(A^0, X)$, relative to the $(e, A)$ transition peak is observed with decreasing thickness, which can be attributed to confinement effects that are amplified in the thinner samples. Looking at the PL spectrum of the 15 nm sample, a clear blue shift in the band edge transition peaks is observed. This is likely due to a combination of the decrease in crystalline defects as well as the aforementioned confinement effects that accompany a reduction in thickness on this scale. The consistency of our PL spectra with the literature suggest these films to be of excellent crystalline quality.

\subsection{\label{sec:level2}Fitting and Removal of Substrate Photoluminescence Peaks}

\begin{figure}
    \centering
    \includegraphics[width=3in]{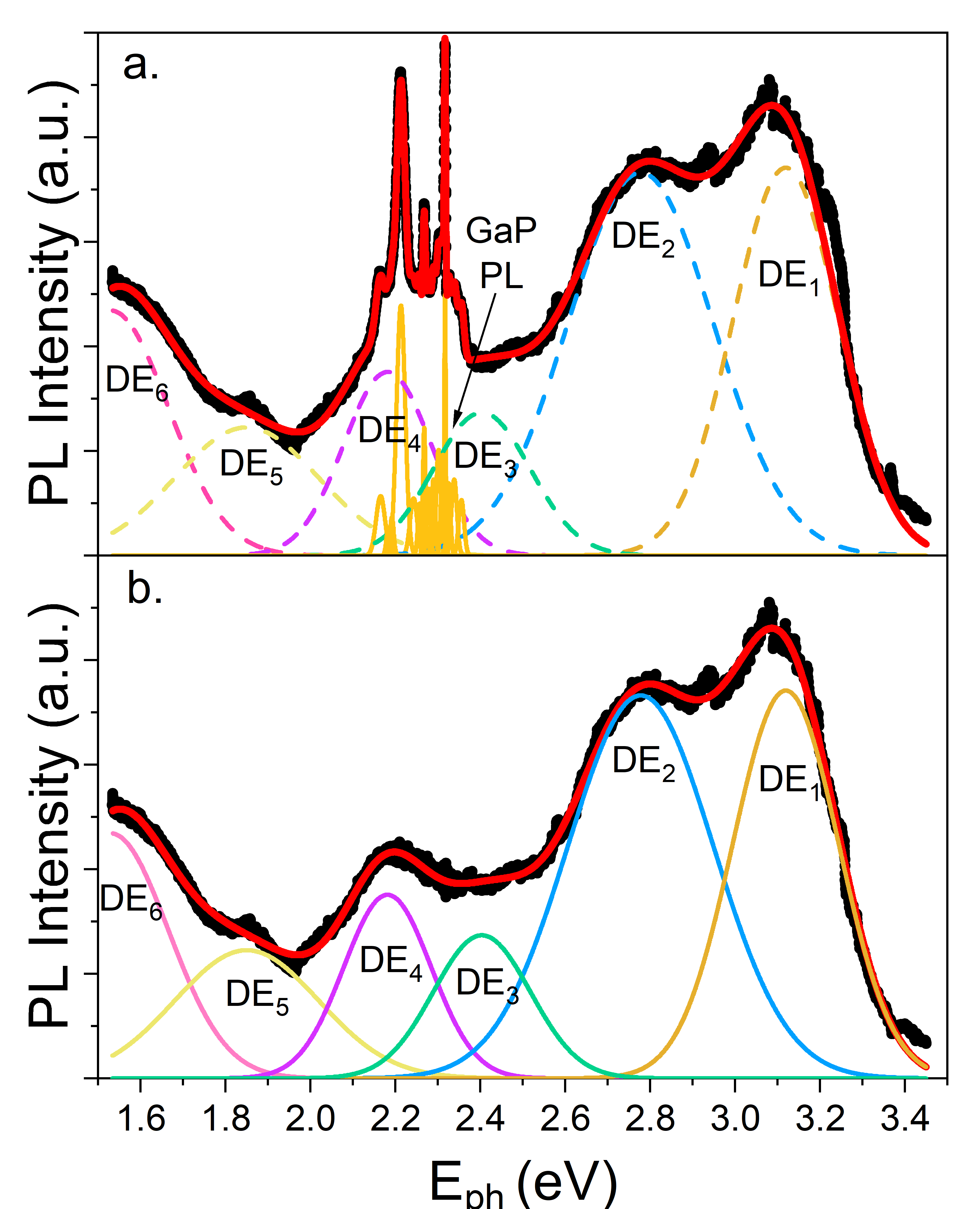}
    \caption{a.) Fitted ZnS deep level emission peaks ($DE_1$-$DE_6$) and GaP band edge PL peaks and b.) same spectrum but with GaP band edge PL peaks removed in post processing. 
    \label{supfig2}
    }
\end{figure}

\begin{figure}
    \centering
    \includegraphics[width=3in]{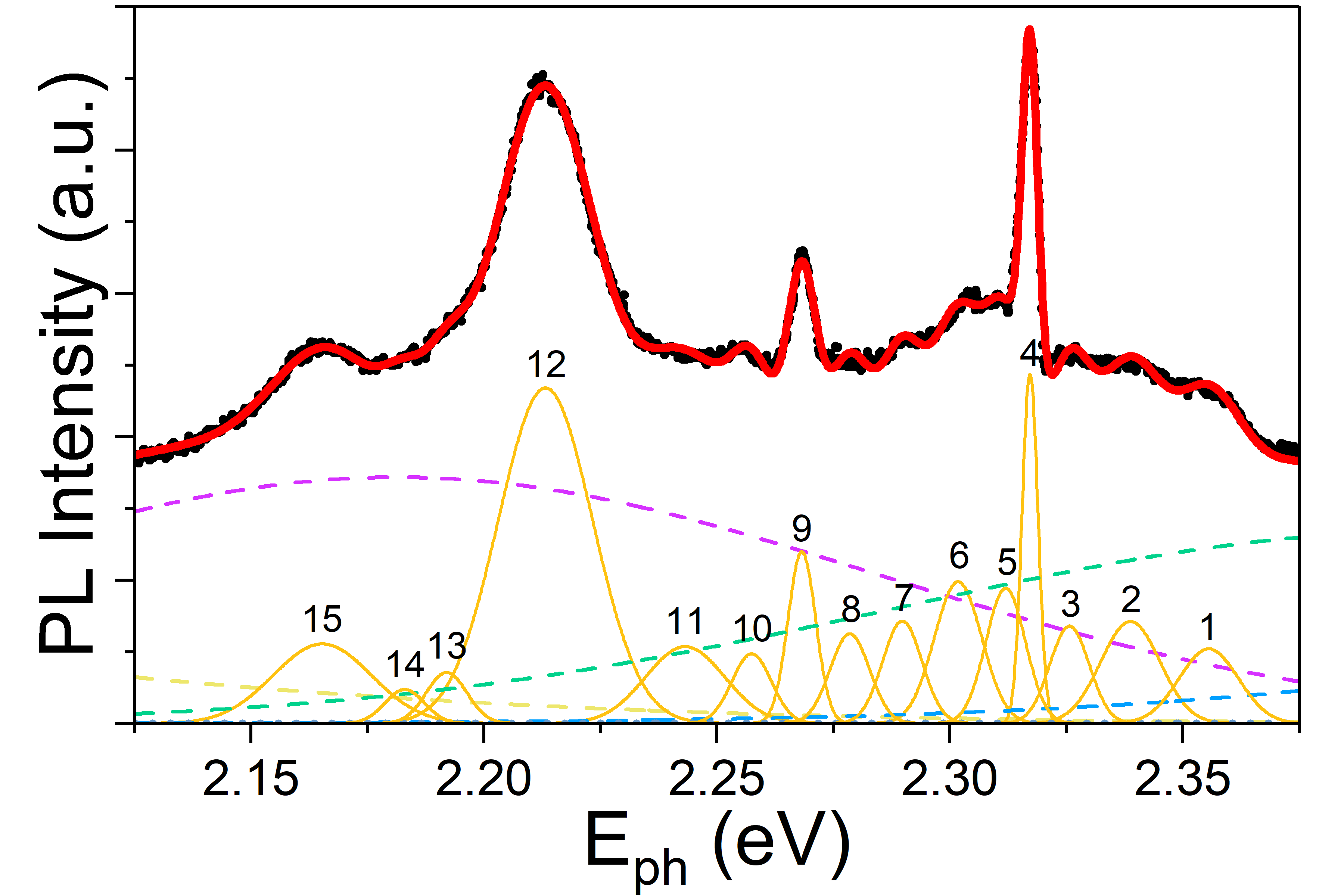}
    \caption{Zoom-in of PL spectrum shown in Fig. 7a to show numbered GaP band edge transition peaks. The comparatively broad ZnS deep level emission peaks are a plotted with dashed line.
    \label{supfig3}
    }
\end{figure}

Due to the thin thickness of the ZnS epilayers, a substantial amount of PL from the GaP substrate is detected between 2.1 eV and 2.4 eV, obscuring two of the ZnS deep level emission peaks, $DE_3$ and $DE_4$ (Fig. \ref{supfig2}a). In order to properly fit $DE_3$ and $DE_4$ and the other ZnS deep level emission peaks, the GaP PL peaks are included in comprehensive fits of deep level emission ranging in energy from 1.5 - 3.5 eV. A close fit is obtained by accounting for PL peaks originating from 15 different GaP band edge transitions (Fig. \ref{supfig3}). The peak energies align closely with previously identified transition energies measured in GaP \cite{alawadhi_indirect_1997, kafi_selective_2024} (Table \ref{suptable2}). In order to eliminate potential confusion for readers, the fitted GaP PL peaks are removed in the main text figures leaving behind only ZnS related PL (Fig. \ref{supfig2}b).

\begin{table*}
\centering
\begin{tabular}{|l||l|l|l|l|}
\hline
\multicolumn{5}{|c|}{GaP Peaks} \\
\hline
Peak Num. & Transition & Energy (eV) – This Study & Energy (eV) – Alawadi et al. & Energy (eV) – Kafi et al. \\
\hline
     
     1 &$E_g(X-\Gamma)+LA$ & 2.356 & 2.361 & N/A\\
     
     2 &$E_g(X-\Gamma)+TA$ & 3.339 & 2.342& 2.340\\
     
     3 &$FX$& 2.326 &  2.326 & N/A\\
     
     4 &$NBX$& 2.317 &  2.3 & 2.318\\
     
     5 &$SBX$& 2.312 &  2.312 & N/A\\
     
     6 &$NBX+TA$& 2.303 & 2.302 & 2.305\\
     
     7 &$NBX+2TA$& 2.290 & 2.290 & 2.291\\
     
     8 &$NBX+3TA$& 2.278 &  N/A & N/A\\
     
     9 &$NBX+LO$& 2.269 & 2.68 & 2.269\\
     
     10 &$NBX+LO+TA$& 2.257 &  2.257 & 2.256\\
     
     11 &$NBX+LO+2TA$& 2.243 & 2.242 & 2.243\\
     
     12 &$NBX+2LO$& 2.213& 2.218 & 2.220 \\
     
     13 &$NBX+2LO+TA$& 2.194& 2.208 & N/A\\
     
     14 &$NBX+2LO+2TA$& 2.185& 2.195 & N/A\\
     
     15 &$NBX+3LO$& 2.165& 2.164 & N/A\\
     \hline
    \end{tabular}
    \caption{Summary of measured GaP band edge transition peaks and comparison to previous literature.
    \label{suptable2}
    }
\end{table*}

\begin{figure}
    \centering
    \includegraphics[width=3in]{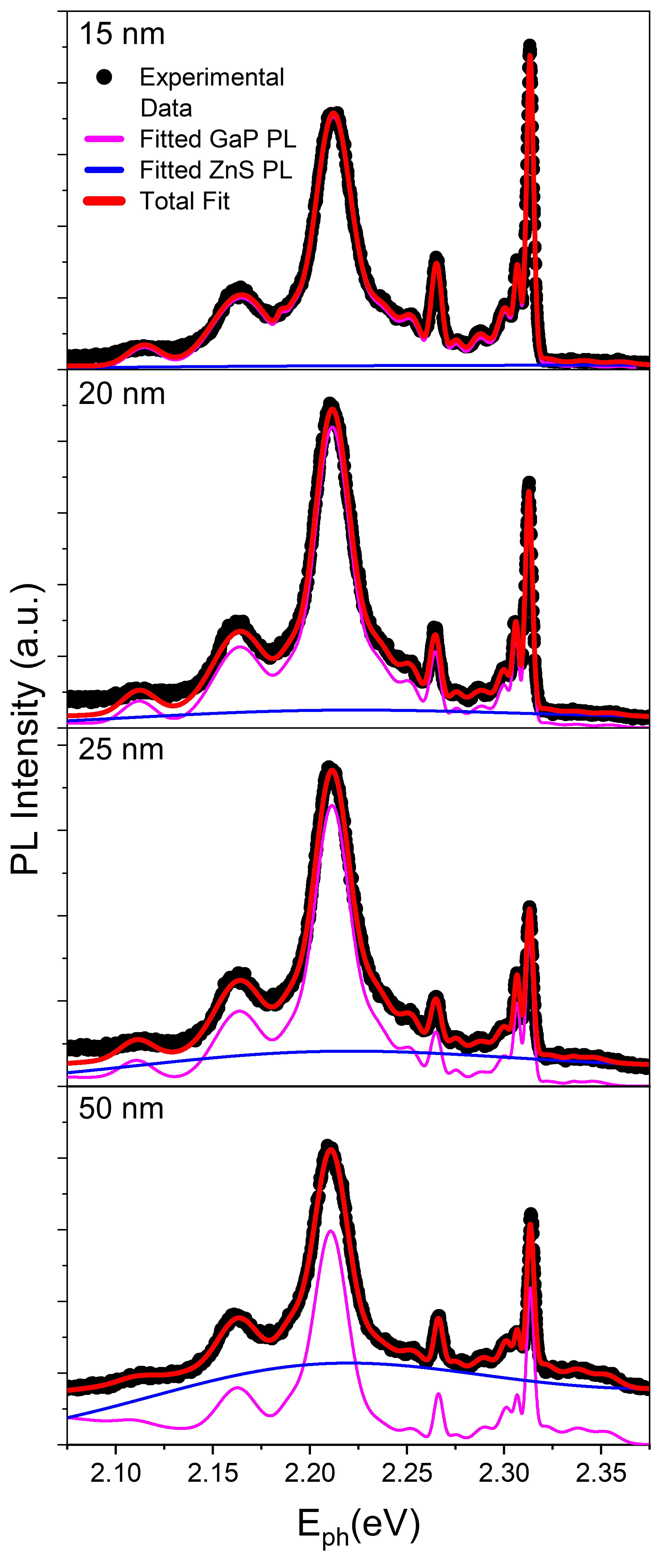}
    \caption{Fitting of GaP PL for the thickness dependent sample set.
    \label{supfig4}
    }
\end{figure}

The GaP peaks are similarly fit for the PL originating from the thickness dependent sample set Fig. \ref{supfig4}). Unsurprisingly, the ZnS epilayer thickness effects the GaP photoluminescence intensity. As epilayer thickness increases the total power from of the excitation source that is able to reach the substrate decreases, thus leading to the observed reduction in GaP PL in the thicker epilayers. Furthermore, PL emitted from the GaP substrate can be potentially absorbed by defect states, including those that we propose to be due to dislocations, further contributing to the reduction in GaP PL at thicker epilayer thicknesses.

\section{\label{sec:level1}Acknowledgements}

Financial support from the Air Force Office of Scientific Research (Grant FA9550-21-1-0278, Program Manager Dr. Ali Sayir) is acknowledged. Support was also provided by the AFOSR Grant No. FA9550-23-1-0330. Computational resources were provided by the Ohio Supercomputer Center.

\bibliography{references}

\end{document}